\documentclass[a4paper]{article}

\usepackage{lscape}
\usepackage{appendix}
\usepackage{amsmath}
\usepackage{amssymb}
\usepackage{amsthm}
\usepackage{txfonts}
\usepackage{graphicx}
\usepackage[all]{xy}
\usepackage[dvipsnames,usenames]{color}
\usepackage{latexsym}
\usepackage{url}

\newtheorem{thrm}{Theorem}

\newtheorem{lemm}[thrm]{Lemma}

\newcommand{\bmpt}[1]{\begin{minipage}[t]{#1\textwidth}}
\newcommand{\bmp}[1]{\begin{minipage}{#1\textwidth}}
\newcommand{\emp}{\end{minipage}}

\newcommand{\cX}{{\cal X}}

\newcommand{\cN}{{\cal N}}

\newcommand{\cO}{{\cal O}}
\newcommand{\bv}{\begin{verbatim}}
\newcommand{\ev}{\end{verbatim}}

\newcommand{\E}{\ensuremath{\mathbb{E}}}
\newcommand{\R}{\ensuremath{\mathbb{R}}}
\newcommand{\ben}{\begin{enumerate}}
\newcommand{\een}{\end{enumerate}}
\newcommand{\be}{\begin{equation}}
\newcommand{\ee}{\end{equation}}
\newcommand{\bea}{\begin{eqnarray}}
\newcommand{\eea}{\end{eqnarray}}
\newcommand{\beaa}{\begin{eqnarray*}}
\newcommand{\eeaa}{\end{eqnarray*}}
\newcommand{\ba}{\begin{align}}
\newcommand{\ea}{\end{align}}
\newcommand{\baa}{\begin{align*}}
\newcommand{\eaa}{\end{align*}}
\newcommand {\bd}{\begin{description}}
\newcommand {\ed}{\end{description}}
\newcommand{\bi}{\begin{itemize}}
\newcommand{\ei}{\end{itemize}}
\newcommand{\bc}{\begin{center}}
\newcommand{\ec}{\end{center}}
\newcommand {\bq}{\begin{quotation}}
\newcommand {\eq}{\end{quotation}} 
\newcommand {\bdo}{\begin{document}} 
\newcommand {\edo}{\end{document}} 
\newcommand{\eql}{\: = \:}
\newcommand{\dfl}{\: \equiv \:}
\newcommand{\onder}[2]{#1_{\mbox{\scriptsize #2}}}
\newcommand {\boven}[2]{#1^{\mbox{\scriptsize #2}}}
\newcommand{\argmin}[1]{\mathop{\rm argmin}\limits_{#1}}
\newcommand{\klein}[2]{#1_{\mbox{\tiny #2}}}
\newcommand{\doo}{\mbox{do}}
\newcommand{\fracd}[2]{{{\displaystyle #1} \over {\displaystyle #2}}}

\newcommand{\intarg}[2]{\!#2\,d#1}
\newcommand{\del}[3]{\left#1 #3 \right#2}
\newcommand{\av}[1]{\del{<}{>}{#1}}
\newcommand{\avv}[1]{\del{<<}{>>}{#1}}
\newcommand{\bra}[1]{\del{<}{|}{#1}}
\newcommand{\ket}[1]{\del{|}{>}{#1}}
\newcommand{\avmini}[1]{\del{<}{>_{\setminus i}}{#1}}
\newcommand{\avminij}[1]{\del{<}{>_{\setminus i,j}}{#1}}
\newcommand{\sminij}{s_{\setminus i,j}}
\newcommand{\ui}{\avmini{u_i^2}}
\newcommand{\hi}{\avmini{h_i}}
\newcommand{\br}[1]{\del{\{}{\}}{#1}}
\newcommand{\bl}[1]{\del{[}{]}{#1}}

\newcommand {\Hxx}{H^{(xx)}}
\newcommand {\Hxy}{H^{(xy)}}
\newcommand {\Hyx}{H^{(yx)}}
\newcommand {\Hyy}{H^{(yy)}}
\newcommand {\Cxx}{C^{(xx)}}
\newcommand {\Cxy}{C^{(xy)}}
\newcommand {\Cyx}{C^{(yx)}}
\newcommand {\Cyy}{C^{(yy)}}
\newcommand {\cx}{c^{(x)}}
\newcommand {\hy}{h^{(y)}}
\newcommand{\bt}{\bar{t}}
\newcommand {\bx}{\bar{x}}
\newcommand {\bX}{\bar{X}}
\newcommand {\by}{\bar{y}}
\newcommand {\bw}{\bar{w}}
\newcommand {\bh}{\bar{h}}
\newcommand {\vs}{\vec{s}}
\newcommand {\vbfs}{{\bf s}}
\newcommand {\vv}{\vec{v}}
\newcommand {\vlambda}{\vec{\lambda}}
\newcommand {\vh}{\vec{h}}
\newcommand {\vk}{\vec{k}}
\newcommand {\vz}{\vec{z}}
\newcommand {\vx}{\vec{x}}
\newcommand {\valpha}{\vec{\alpha}}
\newcommand {\vm}{\vec{m}}
\newcommand {\vmu}{\vec{\mu}}
\newcommand {\vtheta}{\vec{\theta}}
\newcommand {\vgamma}{\vec{\gamma}}
\newcommand {\qd}{q_{\partial_i}}
\newcommand {\qm}{q_{\setminus i}}
\newcommand{\vxv}{\vec{x}_v}
\newcommand{\vxh}{\vec{x}_h}
\newcommand {\vy}{\vec{y}}
\newcommand {\bfy}{{\bf y}}
\newcommand {\bfx}{{\bf x}}
\newcommand {\bft}{{\bf t}}
\newcommand {\bfm}{{\bf m}}
\newcommand {\bfu}{{\bf u}}
\newcommand {\bfw}{{\bf w}}
\newcommand {\bfW}{{\bf W}}
\newcommand {\bfmu}{{\bf \mu}}
\newcommand {\vw}{\vec{w}}
\newcommand {\vb}{\vec{b}}
\newcommand {\vu}{\vec{u}}
\newcommand {\vW}{\vec{W}}
\newcommand {\vM}{\vec{M}}
\newcommand {\sign}{\mbox{sign}}
\newcommand {\sech}{\mbox{sech}}
\newcommand {\hmu}{h^{\mu}_j}
\newcommand {\gmu}{g^{\mu}_j}
\newcommand {\gjkmu}{g^{\mu}_{jk}}
\newcommand {\sia}{s_i^\alpha}
\newcommand {\sjb}{s_j^\beta}
\newcommand {\hia}{h_i^\alpha}
\newcommand {\hjb}{h_j^\beta}
\newcommand {\mia}{m_i^\alpha}
\newcommand {\mib}{m_i^\beta}
\newcommand {\mjb}{m_j^\beta}
\newcommand {\wijab}{w_{ij}^{\alpha\beta}}
\newcommand {\chiijab}{\chi_{ij}^{\alpha\beta}}
\newcommand {\Cijab}{C_{ij}^{\alpha\beta}}
\newcommand {\Hijab}{H_{ij}^{\alpha\beta}}
\newcommand {\he}{h_{\mbox{ext}}}
\newcommand {\xia}{x_{i,\alpha}}
\newcommand {\vxk}{\vec{x}_{\mbox{{\small known}}}}
\newcommand {\vxu}{\vec{x}_{\mbox{{\small unknown}}}}
\newcommand {\Pab}{P_{\alpha\beta}}
\newcommand {\dab}{\delta^{\alpha\beta}}
\newcommand {\dij}{\delta_{ij}}
\newcommand {\erf}{\mbox{Erf}}

\graphicspath{
{/Users/bertkappen/doc/proposals/erc2015/}
}

\title{Adaptive importance sampling for control and inference} 
\author{H.J. Kappen and H. Ruiz
}
\begin{document}
\begin{abstract}
Path integral (PI) control problems are a restricted class of non-linear control problems that can be solved formally as a Feyman-Kac path integral and can be estimated using Monte Carlo sampling.
In this contribution we review path integral control theory in the finite horizon case.

We subsequently focus on the problem how to compute and represent control solutions. Within the PI theory, the question of how to compute becomes the question of importance sampling. Efficient importance samplers are state feedback controllers and the use of these requires an efficient representation. 
Learning and representing effective state-feedback controllers for non-linear stochastic control problems is a very challenging, and largely unsolved, problem. We show how to learn and represent such controllers using ideas from the cross entropy method.
We derive a gradient descent method that 
allows to learn feed-back controllers using an arbitrary parametrisation. We refer to this method as the Path Integral Cross Entropy method or PICE. 
We illustrate this method for some simple examples.

The path integral control methods can be used to estimate the posterior distribution in latent state models. In neuroscience these problems arise when estimating connectivity from neural recording data using EM. We demonstrate the path integral control method as an accurate alternative to particle filtering. 

\end{abstract}
\maketitle
\section{Introduction}
Stochastic optimal control theory (SOC) considers the problem to compute an optimal sequence of actions to attain a future goal. 
The optimal control is usually computed from the Bellman equation, which is a partial differential equation. Solving the equation for high dimensional systems
is difficult in general, except for special cases, most notably the case of linear dynamics and  quadratic control cost or the noiseless deterministic case. Therefore, despite its elegance
and generality, SOC has not been used much in practice.

In \cite{fleming1982optimal} it was observed that posterior inference in a certain class of diffusion processes can
be mapped onto a stochastic optimal control problem. These so-called
Path integral (PI) control problems \cite{kap05a} represent a restricted class of non-linear control problems with arbitrary dynamics and state cost, but with a linear dependence of the control on the dynamics and quadratic control cost. For this class of control problems, the Bellman equation can be transformed into a linear partial differential equation.
The solution for both the optimal control and the optimal cost-to-go can be expressed in closed form as a Feyman-Kac path integral. The path integral involves an expectation value with respect to a dynamical system. As a result, the optimal control can be estimated using Monte Carlo sampling. See \cite{todorov_pnas2009,kappen2009c,kappen-gomez-opper-mlj2009} for earlier reviews and references. 

In this contribution we review path integral control theory in the finite horizon case.
Important questions are: how to compute and represent the optimal control solution. In order to efficiently compute, or approximate, the optimal control solution we discuss the notion of importance sampling and the relation to the Girsanov change of measure theory. As a result, the path integrals can be estimated using (suboptimal) controls. Different importance samplers all yield the same asymptotic result, but differ in their efficiency. We show an intimate relation between optimal importance sampling and optimal control: we prove a Lemma that shows that the optimal control solution {\em is} the optimal sampler, and better samplers (in terms of effective sample size) are better controllers (in terms of control cost) \cite{thijssen2014a}. 
This allows us to iteratively improve the importance sampling, thus increasing the efficiency of the sampling.

In addition to the computational problem, another key problem is the fact that the optimal control solution is in general a state- and time-dependent function $u(x,t)$ with $u$ the control, $x$ the state and $t$ the time. The state dependence is referred to as a feed-back controller, which means that the execution of the control at time $t$ requires knowledge of the current state $x$ of the system. It is often impossible to compute the optimal control for all states because this function is an infinite dimensional object, which we call the {\em representation problem}. 
Within the robotics and control community, there are several approaches to deal with this problem. 

\subsection*{Deterministic control and local linearisation}
The simplest approach follows from the realisation that state-dependent control is only required due to the noise in the problem. In the deterministic case, one can compute the optimal control solution $u(t)=u^*(x^*(t),t)$ along the optimal path $x^*(t)$ only, and this is a function that only depends on time. This is a so-called open loop controller which applies the control $u(t)$ regardless of the actual state that the system is at time $t$. This approach works for certain robotics tasks such a grasping or reaching. See for instance  \cite{Theodorou_JMLR2010,schaal2010} who constructed open loop controllers for a number of robotics tasks within the path integral control framework. However, open loop controllers are clearly sub-optimal in general and simply fail for unstable systems that require state feedback. 

It should be mentioned that the open loop approach can be stabilised by computing a linear feed-back controller {\em around} the deterministic trajectory. This approach uses the fact that for linear dynamical systems with Gaussian noise and with quadratic control cost, the solution can be efficiently computed. \footnote{For these so-called linear quadric control problems (LQG) the optimal cost-to-go is quadratic in the state and the optimal control is linear in the state, both with time dependent coefficients. The Bellman equation reduces to a system of non-linear ordinary differential equations for these coefficients, known as the Ricatti equation.} One defines a linear quadratic control problem around the deterministic optimal trajectory $x^*(t)$ by Taylor expansion to second order, which can be solved efficiently. The result is a linear feedback controller that stabilises the trajectory $x^*(t)$. This two-step approach is well-known and powerful and at the basis of many control solutions such as the control of ballistic missiles or chemical plants \cite{stengel93}. 

The solution of the linear quadratic control problem also provides a correction to the optimal trajectory $x^*(t)$. Thus, a new $x^*(t)$ is obtained and a new LGQ problem can be defined and solved.
This approach can be iterated, incrementally improving the trajectory $x^*(t$) and the linear feedback controller. This approach is known as Differential Dynamic Programming \cite{mayne1966,murray1984differential} or the Iterative LQG method \cite{todorov2005}. In the robotics community this is a popular method, providing a practical compromise between stability, non-linearity and efficient computation \cite{morimoto2003minimax,tassa2011theory, tassa2014control}.

\subsection*{Model predictive control}
The second approach is to compute the control 'at run-time' for any state that is visited using the idea of model predictive control (MPC) \cite{camacho2013model}. At each time $t$ in state $x_t$, one defines a finite horizon control problem on the interval $[t,t+T]$ and computes the optimal control solution $u(s,x_s), t\le s\le t+T$ on the entire interval. One executes the dynamics using  $u(t,x_t)$ and the system moves to a new state $x_{t+dt}$ as a result of this control and possible external disturbances. This approach is repeated for each time. The method relies on a model of the plant and external disturbances, and on the possibility to compute the control solution sufficiently fast. 
MPC yields a state dependent controller because the control solution in the future time interval depends on the current state. MPC avoids  the representation problem altogether, because the control is never explicitly represented for all states, but computed for any state when needed. MPC is particularly well-suited for the path integral control problems, because in this case the optimal control $u^*(x,t)$ is explicitly given in terms of a path integral. The challenge then is to evaluate this path integral sufficiently accurate in real time. \cite{thijssen2014a} propose adaptive Monte Carlo sampling that is accelerated using importance sampling.  This approach has been successfully applied to the control of 10 to 20 autonomous helicopters (quadrotors) that are engaged in coordinated control tasks such as flying with minimal velocity in a restricted area without collision or a task where multiple 'cats' need to catch a mouse that tries to get away \cite{gomez2015real}.

\subsection*{Parametrized solution}
The third approach is to consider a parametrised family of controllers $u(t,x||\theta)$ and to find the optimal parameters $\theta^*$. If successful, this yields a near optimal state feedback controller for all $t,x$. This approach is well-known in the control and reinforcement community. Reinforcement learning (RL) is a particular setting of control problems with the emphasis on learning a controller on the basis of trial-and-error. A sequence of states $X_t, t=0,dt,2dt,\ldots,$ is generated from a single roll-out of the dynamical system using a particular control, which is called the policy in RL. The 'learning' in reinforcement learning refers to the estimation of the optimal policy or cost-to-go function from a single roll out \cite{sutton98}. 
The use of function approximation in RL is not straightforward \cite{bellman1959functional,sutton88,bertsekas-tsitsiklis1996}. To illustrate the problem, consider the infinite horizon discounted reward case, which is the most popular RL setting. The problem is to compute the optimal cost-to-go of a particular parametrised form: $J(x|\theta)$. In the non-parametrised case, the solution is given by the Bellman 'back-up' equation, which relates $J(x_t)$ to $J(x_{t+dt})$ where $x_{t,t+dt}$ are the states of the system at time $t,t+dt$, respectively and $x_{t+dt}$ is related to $x_t$ through the dynamics of the system. In the parametrised case, one must compute the new parameters $\theta'$ of $J(x_t|\theta')$ from $J(x_{t+dt}|\theta)$ . The problem is that  the update is 
in general not of the parametrised form and an additional approximation is required to find the $\theta'$ that gives the best approximation.
In the RL literature, one makes the distinction between 'on-policy' learning where $J$ is only updated for the sequence of states that are visited, and off-policy learning updates $J(x)$ for all states $x$, or a (weighted) set of states. Convergence of RL with function approximation has been shown for on-policy learning with linear function approximation (ie. $J$ is a linear function of $\theta$) \cite{tsitsiklis1997analysis}. These authors also provide examples of both off-policy learning and non-linear function approximation where learning does not converge.

\subsection*{Outline}

This chapter is organized as follows. In section~\ref{section:setup} we present a review of the main ingredients of the path integral control method. We define the path integral control problem and state the basic Theorem of its solution in terms of a path integral. We then prove the Theorem by showing in section~\ref{section:hjb} that the Bellman equation can be linearized by a log transform and in section~\ref{section:ito} that the solution of this equation is given in terms of a Feyman-Kac path integral. In section~\ref{section:mc} we discuss how to efficiently estimate the path integral using the idea of importance sampling. We show that the optimal importance sampler coincides with the optimal control.  In section~\ref{section:ce} we review the cross entropy method, as an adaptive procedure to compute an optimized importance sampler in a parametrized family of distributions. In order to apply the cross entropy method, we reformulate the path integral control problem in terms of a KL divergence minimization in section~\ref{section:kl} and in section~\ref{section:ce+pi} we apply this procedure to the obtain optimal samplers/controllers to estimate the path integrals. In section~\ref{section:numerical} we illustrate the method to learn a parametrized time independent state dependent controller for some simple control tasks.

In section~\ref{neural} we consider the reverse connection between
control and sampling: We consider the problem to compute the posterior
distribution of a latent state model that we wish to approximate using
Monte Carlo sampling, and to use optimal controls to accelerate this
sampling problem. 
In neuroscience, such problems arise, e.g. to estimate network connectivity from data or decoding of
neural recordings. The common approach is to formulate a maximum likelihood problem that is optimized using the EM method.
The E-step is a Bayesian inference problem over hidden states and
is shown to be equivalent to a path integral control problem. 
We illustrate this for a small toy neural network where we estimate the neural activity from noisy observations.

\section{Path integral control}\label{section:setup}
Consider the dynamical system 
\begin{align}
&dX(s)=f(s,X(s))ds+g(s,X(s))\Big(u(s,X(s))ds+ dW(s)\Big)\qquad t\leq s\leq T \label{importance}
\end{align}
with $X(t)=x$. 
$dW(s)$ is Gaussian noise with $\E\ dW(s)=0, \E \ dW(s)dW(r) =ds\delta(s-r)$. The stochastic process $W(s), t\le s\le T$ is called a Brownian motion.
We will use upper case for stochastic variables and lower case for deterministic variables. $t$ denotes the current time and $T$ the future horizon time. 

Given a  function $u(s,x)$ that defines the control for each state $x$ and each time $t\le s\le T$, define the cost
\begin{align}
S(t,x,u)&=\Phi(X(T))+\int_t^T\left(V(s,X(s))+\frac12 u(s,X(s))^2\right)ds \nonumber\\
&+\int_t^T u(s,X(s)) dW(s) \label{S}
\end{align}
with $t,x$ the current time and state and $u$ the control function.
The stochastic optimal control problem is to find the optimal control function $u$: 
\begin{align}
J(t,x)&=\min_{u}\E \ S(t,x,u)\nonumber \\
u^*(t,x)&=\arg\min_{u}\E \ S(t,x,u) \label{controlproblem1}
\end{align}where $\E$ is an expectation value with respect to the stochastic process Eq.~\ref{importance} with initial condition $X_t=x$ and control $u$. 

$J(t,x)$ is called the optimal cost-to-go as it specifies the optimal cost from any intermediate state and any intermediate time until the end time $t=T$. For any control problem, $J$ satisfies a partial differential equation known as the Hamilton-Jacobi-Bellman equation (HJB). In the special case of the path integral control problems the solution is given explicitly as follows.
\begin{thrm}\label{thm1}
The solution of the control problem Eqs.~\ref{controlproblem1} is given by 
\bea
J(t,x)&=& -\log \psi(t,x) \qquad \psi(t,x)= \E\ e^{-S(t,x,u)}\label{eq1}\\
u^*(t,x) &=& u(t,x) +\av{\frac{dW(t)}{dt}}\label{eq2}
\eea
where we define
\begin{align}
\av{\frac{dW}{dt}}&=\lim_{s\downarrow t} \frac{1}{s-t}
	\frac{\E\left[W(s) e^{-S(t, x,u)}\right]}
	{\E \left[e^{-S(t, x,u)}\right]}\label{def_brackets}
	\end{align}
and $W(s), s\ge t$ the Brownian motion.
\end{thrm}

The path integral control problem and Theorem~\ref{thm1} can be generalised to 
 the multi-dimensional case where
$X(t), f(s,X(s))$ are $n$-dimensional vectors, $u(s,X(s))$ is an $m$ dimensional vector and $g(s,X(s))$ is an $n\times m$ matrix. 
$dW(s)$ is $m$-dimensional Gaussian noise with $\E\ dW(s)=0$ and $\E \ dW(s)dW(r) =\nu ds\delta(s-r)$ and $\nu$ the $m\times m$ positive definite covariance matrix. Eqs.~\ref{importance} and~\ref{S} become:
\begin{align}
&dX(s)=f(s,X(s))ds+g(s,X(s))\Big(u(s,X(s))ds+ dW(s)\Big)\qquad t\leq s\leq T\nonumber\\
&S(t,x,u)=\frac{1}{\lambda}\left(\Phi(X(T))+\int_t^T\left(V(s,X(s))+\frac12 u(s,X(s))' R u(s,X(s))\right)ds \right.\nonumber\\
&\left.+\int_t^T u(s,X(s))'  R dW(s)\right)\label{ndim}
\end{align}
where $'$ denotes transpose. In this case, $\nu$ and $R$ must be related as 
with $\lambda I = R \nu$ with $\lambda>0$ a scalar \cite{kap05a}.
%

In order to understand this result, we first will derive in section~\ref{section:hjb} the HJB equation and show that for the path integral control problem it can be transformed into a linear partial differential equation. 
Subsequently, in section~\ref{section:ito} we present a Lemma that will allow us prove the Theorem.

\subsection{The linear HJB equation}
\label{section:hjb}
The derivation of the HJB equation relies on the argument of dynamic programming. This is quite general, but here we restrict ourselves to the path integral case.
Dynamic programming
expresses the control problem on the time interval $[t,T]$ as an instantaneous contribution at the small time interval $[t,t+ds]$ and a control problem on the interval $[t+ds,T]$. From the definition of $J$ we obtain that $J(T,x)=\Phi(x), \forall x$.

We derive the HJB equation by discretising time with infinitesimal time increments $ds$. The dynamics and cost-to-go become
\beaa
x_{s+ds}&=&x_s +f_s(X_s)ds+g_s(X_s)\Big(u_s(X_s)ds+ dW_s\Big) \qquad s=t,t+ds,\ldots,T-ds\\
S_t(x,u_{t:T-ds})&=&\Phi(x_T)+\sum_{s=t}^{T-ds}ds \left(V_s(X_s)+\frac12 u_s(X_s)^2\right) +\sum_{s=t}^{T-ds} u_s(X_s) dW_s
\label{discretedynamics}
\eeaa
The minimisation in Eq.~\ref{controlproblem1} is with respect to a functions $u$ of state and time and becomes a minimisation over a sequence of state-dependent functions $u_{t:T-ds}=\{u_s(x_s), s=t,t+ds,\ldots, t+T-ds\}$: 
\beaa
J_t(x_t)&=&\min_{u_{t:T-ds}}\E \  S_t(x_t,u_{t:T-ds})\\
&=&\min_{u_{t}} \left(V_t(x_t)ds+\frac12 u_t(x_t)^2 ds+\min_{u_{t+ds:T-ds}} \E \ S_{t+ds}(X_{t+ds}, u_{t+ds:T-ds})\right)\\
&=&\min_{u_t} \left(V_t(x_t)ds+\frac12 u_t(x_t)^2ds+\E \ J_{t+ds}(X_{t+ds})\right)\\
&=&\min_{u_t} \left(V_t(x_t)ds+\frac12 u_t(x_t)^2 ds+J_t(x_t) +ds( f_t(x_t)+g_t(x_t) u_t(x_t))\partial_x J_t(x_t)\right.\\
&+&\left. \frac{1}{2}ds \partial^2_x J_t(x_t)+\partial_t J_t(x_t)ds +\cO(ds^2)\right)\nonumber
\eeaa
The first step is the definition of $J_t$. The second step separates the cost term at time $t$ from the rest of the contributions in $S_t$, uses that $\E dW_t=0$. 
The third step identifies the second term as the optimal cost-to-go from time $t+ds$ in state $X_{t+ds}$. The expectation is with respect to the next future state $X_{t+ds}$ only.  The fourth step uses the dynamics of $x$ to express $X_{t+ds}$ in terms of $x_t$, a first order Tayler expansion in $ds$ and a second order Taylor expansion in $(X_{t+ds}-x_t$ and uses the fact that $\E X_{t+ds}-x_t =(f_t(x_t) + g_t(x_t)u_t(x_t))ds$ and $\E (X_{t+ds}-x_t)^2 = \E dW_t^2 +\cO(ds^2)=ds +\cO(ds^2)$.
$\partial_{t,x}$ are partial derivatives with respect to $t,x$ respectively.

Note, that the minimization of control paths $u_{t:T-ds}$ is absent in the final result, and only a minimization over $u_t$ remains. 
We obtain in the limit $ds\to 0$:
\bea
-\partial_t J(t,x)&=&\min_u\left(V(t,x)+\frac{1}{2}u^2(t,x) + (f(t,x)+g(t,x)u(t,x))\partial_x
J(x,t)\right.\nonumber\\
&+&\left.\frac{1}{2}g(t,x)^2 \partial_x^2 J(t,x)\right)\label{hjbe}
\eea
Eq.~\ref{hjbe} is a partial differential equation,
known as the {\em Hamilton-Jacobi-Bellman (HJB) equation}, 
that describes the evolution of $J$ as a function of $x$ and $t$ and must
be solved with boundary condition $J(x,T)=\phi(x)$. 

Since $u$ appears linear and quadratic in Eq.~\ref{hjbe}, we can solve the minimization with respect to $u$ which gives
$u^*(t,x)=-  g(t,x) \partial_x J(t,x)$. 
Define $\psi(t,x)=e^{-J(t,x)}$,  then the HJB equation becomes linear in $\psi$:
\bea
\partial_t \psi + f \partial_x \psi + \frac{1}{2}  g^2 \partial_x^2 \psi=  V\psi.\label{linear_hjb}
\eea
with boundary condition $\psi(T,x)=e^{-\Phi(x)}$.

\subsection{Proof of the Theorem}
\label{section:ito}
In this section we show that Eq.~\ref{linear_hjb} has a solution in terms of a path integral (see \cite{thijssen2014a}). 
In order to prove this, we first derive the following Lemma. The derivation makes use of the so-called It\^{o} calculus which we have summarised in the appendix.

\begin{lemm}
\label{lemma} Define the stochastic processes $Y(s),Z(s), t\le s\le T$ as functions of the stochastic process Eq.~\ref{importance}:
\bea
Z(s)=\exp(-Y(s)))\quad Y(s)=\int_t^s V(r,X_r)dr + \frac{1}{2} u(r,X_r)^2dr+u(r,X_r) dW(r)\quad t\le s\le T\label{Z}
\eea
When $\psi$ is a solution of the linear Bellman equation Eq.~\ref{linear_hjb} and $u^*$ is the optimal control, then
\bea
e^{-S(t,x,u)} - \psi(t,x) &= \int_{t}^{T} Z(s)\psi(s,X_s) (u^*(s,X_s)-u(s,X_s)) dW(s) \label{eq:pi}
\eea
 \end{lemm}

\begin{proof}
Consider $\psi(s,X(s)), t\le s\le T$ as a function of the stochastic process Eq.~\ref{importance}.
Since $X(s)$ evolves according to Eq.~\ref{importance}, $\psi$ is also a stochastic process and 
we can use  
It\^o's Lemma (Eq.~\ref{ito_lemma} to derive a dynamics for $\psi$. 
\begin{align*}
&d\psi
	=\left( \partial_t \psi + (f+gu) \partial_x \psi + \frac12  \ g^2 \partial_x^2 \psi\right)ds
		+g dW \partial_x \psi =V\psi ds + g(uds+dW)\partial_x \psi
\end{align*}
where the last equation follows because $\psi$ satisfies the linear Bellman equation Eq.~\ref{linear_hjb}.

From the definition of $Y$ we obtain $dY=Vds + \frac{1}{2}u^2 ds + u dW$. 
Using again It\^o's Lemma Eq.~\ref{ito_lemma}:
\beaa
dZ =-Z dY + \frac{1}{2}Z d[Y,Y]= -Z\left( V ds +udW\right)
\eeaa
Using the product rule Eq.~\ref{ito_product} we get
\beaa
d(Z\psi)=\psi dZ+Z d\psi+d[Z,\psi]=  -Z\psi u dW+Z\partial_x \psi g dW=Z\psi(u^*-u)dW\label{dzpsi}
\eeaa
where in the last step we used that $u^*=\frac{1}{\psi} g \partial_x \psi$ which follows from $u^*(t,x)=-  g(t,x) \partial_x J(t,x)$. 
and $\psi(t,x)=e^{-J(t,x)}$ (see section~\ref{section:hjb}). 
Integrating $d(Z\psi)$ from $t$ to $T$ using Eq.~\ref{ito_integral} yields
\beaa
Z(T)\psi(T)-Z(t)\psi(t,x)&=&\int_t^T d(Z\psi)\\
e^{-Y(T)-\Phi(X(T))}-\psi(t,x)&=&\int_t^T ds Z\psi(u^*-u)dW
\eeaa
where we used that $Z(t)=1$ and $\psi(T)=\exp(-\Phi(X(T)))$. This proves Eq.~\ref{eq:pi}.
\end{proof}

With the Lemma, it is easy to prove Theorem~\ref{thm1}. Taking the expected value in Eq.~\ref{eq:pi} proves Eq.~\ref{eq1}
\begin{align*}
\psi(t,x) = \E\left[e^{-S(t, x,u)}\right]
\end{align*}
This is a closed form expression for the optimal cost-to-go as a path integral.

To prove Eq.~\ref{eq2}, we multiply  Eq.~\ref{eq:pi} with $W(s)=\int_{t}^{s}dW$, which is an increment of the Wiener Process and take the expectation value:
\begin{align*}
\E\left[
	e^{-S(t,x,u)}
	W(s)
\right]
=\E\left[
\int_t^sZ\psi (u^*-u)  dW
\int_t^sdW
\right]=\int_t^s\E\left[
Z\psi(u^*-u) \right] dr
\end{align*}
where in the first step we used $\E W(s)=0$ and in the last step we used  It\^o Isometry Eq.~\ref{ito_iso}.
To get $u^*$ we divide by the time increment $s-t$ and take the limit of the time increment to zero. This will yield the integrand of the RHS $\psi(t,x)(u^*(t,x)-u(t,x)$. Therefore the expected value disappears and we get
\begin{align*}
u^*(t,x)=u(t,x)+\frac{1}{\psi(t,x)} \lim_{s\downarrow t}\frac1{s-t}
\E\left[
	e^{-S(t,x,u)}
	W(s)
\right] 
\end{align*}
which is Eq.~\ref{eq2}.

\subsection{Monte Carlo sampling}
\label{section:mc}
Theorem~\ref{thm1} gives an explicit expression for the optimal control $u^*(t,x)$ and the optimal cost-to-go $J(t,x)$  in terms of an expectation value over trajectories that start at $x$ at time $t$ until the horizon time $T$. One can estimate the expectation value by Monte Carlo sampling. One generates $N$ trajectories $X(t)_i, i=1,\ldots, N$ starting at $x,t$ that evolve according to the dynamics Eq.~\ref{importance}. 
Then, $\psi(t,x)$ and $u^*(t,x)$ are estimated as
\bea
\hat{\psi}(t,x)&=&\sum_{i=1}^N w_i \qquad  w_i= \frac{1}{N} e^{-S_i(t,x,u)}
\label{mc_psi1}\\
\hat{u}^*(t,x)&=&u(t,x)+\frac{1}{\hat{\psi}(t,x)}\lim_{s\downarrow t} \frac{1}{s-t}\sum_{i=1}^N W(s)_i w_i
\eea
with $S_i(t,x,u)$ the value of $S(t,x,u)$ from Eq.~\ref{S} for the $i$th trajectory $X(s)_i, W(s)_i, t\le s\le T$.
The optimal control estimate involves a limit which we must handle numerically by setting $s-t=\epsilon>0$. Although in theory the result holds in the limit $\epsilon\to 0$, in practice $\epsilon$ should be taken a finite value because of numerical instability, at the expense of theoretical correctness. 

The estimate involves a control $u$, which we refer to as the sampling control. Theorem~\ref{thm1} shows that one can use {\em any} sampling control to compute these expectation values. 
The choice of $u$ affects the efficiency of the sampling. The efficiency of the sampler depends on the variance of the weights $w_i$ which can be easily understood. If the weight of one sample dominates all other weights, the weighted sum over $N$ terms is effectively only one term. The optimal weight distributions for samping is obtained when all samples contribute equally, which means that all weights are equal. It can be easily seen from Lemma~\ref{lemma} that this is obtained when $u=u^*$. In that case, the right hand side of Eq.~\ref{eq:pi} is zero and thus is $S(t,x,u^*)$ a deterministic quantity. 
This means that for all trajectories $X_i(t)$ the value $S_i(t,x,u^*)$ is the same (and equal to the optimal cost-to-go $J(t,x)$). Thus, sampling with $u^*$ has {\em zero variance} meaning that all samples yield the same result and therefore only one sample is required. 

One can view the choice of $u$ as implementing a type of importance sampling and
the optimal control $u^*$ {\em is} the optimal importance sampler. One can also deduce from Lemma~\ref{lemma} that when $u$ is close to $u^*$, the variance in the right hand side of Eq.~\ref{eq:pi} as a result of the different trajectories is small and thus is the variance in $w_i=e^{-S_i(t,x,u)}$ is small. Thus, the closes $u$ is to $u^*$ the more effective is the importance sampler \cite{thijssen2014a}. 

Since it is in general not feasible to compute $u^*$ exactly, the key question is how to compute a good approximation to $u^*$. In order to address this question, we propose the so-called cross-entropy method.

\section{The cross-entropy method}
\label{section:ce}
The cross-entropy method \cite{de2005tutorial} is an adaptive approach to importance sampling. 
Let $X $ be a random variable taking values in the space $\cX$. 
Let $f_v(x)$ be a family of probability density function on $\cX$ parametrized by $v$ and $h(x)$ be a positive function. Suppose that we are interested in the expectation value \bea
l=\E_u \ h =\int dx f_u(x) h(x) \label{ce2}
\eea
where $\E_u$ denotes expectation with respect to the pdf $f_u$ for a particular value of $v=u$.
A crude estimate of $l$ is by naive Monte Carlo sampling from $f_u$: Draw $N$ samples $X_i,i=1,\ldots, N$ from $f_u$ and construct the estimator
\bea
\hat{l}=\frac{1}{N}\sum_{i=1}^N h(X_i)\label{ce5}
\eea
The estimator is a stochastic variable and is unbiased, which means that its expectation value is the quantity of interest: $\E_u \hat{l}=l$. The variance of $\hat{l}$ quantifies the accuracy of the sampler. 
The accuracy is high when many samples give a significant contribution to the sum. However, when the supports of $f_u$ and $h$ have only a small overlap, most samples $X_i$ from $f_u$ will have $h(X_i)\approx 0$ and only few samples effectively contribute to the sum. In this case the estimator has high variance and is inaccurate. 

A better estimate is obtained by {\em importance sampling}. The idea is to define an importance sampling distribution $g(x)$ and to sample $N$ samples from $g(x)$ and construct the estimator:
\bea
\hat{l}=\frac{1}{N}\sum_{i=1}^N h(X_i) \frac{f_u(X_i)}{g(X_i)}\label{ce6}
\eea
It is easy to see that this estimator is also unbiased: $\E_g \hat{l} = \frac{1}{N}\sum_i \E_g  h(X)\frac{f_u(X)}{g(X)}=\E_u h(X) = l$. The question now is to find a $g$ such that $\hat{l}$ has low variance. When $g=f_u$ Eq.~\ref{ce6} reduces to Eq.~\ref{ce5}.

Before we address this question, note that it is easy to construct the optimal importance sampler. It is given by 
\beaa
g^*(x)=\frac{h(x) f_u(x)}{l}
\eeaa
where the denominator follows from normalization: $1=\int dx g^*(x)$. In this case the estimator  Eq.~\ref{ce6} becomes $\hat{l}=l$ for any set of samples. 
Thus, the optimal importance sampler has zero variance and $l$ can be estimated with one sample only. 
Clearly $g^*$ cannot be used in practice since it requires $l$, which is the quantity that we want to compute! 

However, we may find an importance sampler that is close to $g^*$. The cross entropy method suggests to find the distribution $f_v$ in the parametrized family of distributions that minimises the KL divergence
\bea
KL(g^*|f_v)=\int dx g^*(x) \log\frac{g^*(x)}{f_v(x)}\propto -\E_{g^*} \log f_v(X) \propto -\E_u h(X) \log f_v(X) = -D(v) \label{ce3}
\eea
where in the first step we have dropped the constant term $\E_{g^*} \log g^*(X)$ and in the second step have used the definition of $g^*$ and dropped the constant factor $1/l$. 

The objective is to maximize $D(v)$ with respect to $v$. For this we need to compute $D(v)$ which involves an expectation with respect to the distribution $f_u$. We can use again importance sampling to compute this expectation value. Instead of $f_u$ we sample from $f_w$ for some $w$. We thus obtain
\beaa
D(v)=\E_{w} h(X) \frac{f_u(X)}{f_w(X)}\log f_v(X)
\eeaa
We estimate the expectation value by drawing $N$ samples from $f_w$. If $D$ is convex and differentiable with respect to $v$, the optimal $v$ is given by
\bea
\frac{1}{N}\sum_{i=1}^N h(X_i) \frac{f_u(X_i)}{f_w(X_i)}\frac{d}{dv} \log f_v(X_i)=0\qquad X_i \sim f_w\label{ce1}
\eea
The cross entropy method considers the following iteration scheme. Initialize $w_0=u$. In iteration $n=0,1,\ldots$ generate $N$ samples from $f_{w_n}$ and compute $v$ by solving Eq.~\ref{ce1}. Set $w_{n+1}=v$. 

We illustrate the cross entropy method for a simple example. Consider $\cX=\R$ and the family of so-called tilted distributions $f_v(x)=\frac{1}{N_v} p(x)e^{vx}$, with $p(x)$ a given distribution and $N_v=\int dx p(x)e^{vx}$ the normalization constant. We assume that it is easy to sample from $f_v$ for any value of $v$. Choose $u=0$, then the objective  Eq.~\ref{ce2} is to compute $l=\int dx p(x) h(x)$. We wish to estimate $l$ as efficient as possible by optimizing $v$. Eq.~\ref{ce1} becomes
\beaa
\frac{\partial \log N_v}{\partial v}=\frac{ \sum_{i=1}^N h(X_i)e^{-w X_i}X_i }{\sum_{i=1}^N h(X_i) e^{-w X_i} }
\eeaa
Note that the left hand side is equal to $\E_v X$ and the right hand side is the '$h$ weighted' expected $X$ under $p$. The cross entropy update is to find $v$ such that $h$-weighted expected $X$ equals $\E_v X$. This idea is known as moment matching: one finds $v$ such that the moments of the left and right hand side, in this case only the first moment, are equal.

\subsection{The Kullback-Leibler formulation of the path integral control problem}
\label{section:kl}
In order to apply the  cross entropy method to the path integral control theory, we reformulate the control problem Eq.~\ref{importance} in terms of a KL divergence.
Let $\cX$ denote the space of continuous trajectories on the  interval $[t,T]$: $\tau=X(s),t\le s \le T$ with fixed initial value $X(t)=x$.  Denote $p_u(\tau)$ the distribution over trajectories $\tau$ with control $u$.

The distributions $p_u$ for different $u$ are related to each other by the Girsanov Theorem. We derive this relation by simply discretising time as before. In the limit $ds\to 0$, the conditional probability of $X_{s+ds}$ given $X_s$ is Gaussian with mean $\mu_s=X_s+f(s,X_s)ds+g(s,X_s)u(s,x_s)ds$ and variance $\Xi_sds=g(s,X_s)^2 ds$. Therefore, the conditional probability of a trajectory $\tau=X_{t:T}|x$ with initial state $X_t=x$ is
\footnote{In the multi-dimensional case of Eq.~\ref{ndim} this generalizes as follows.  The variance is $g(s,X_s) \nu g(s,X_s)'ds=\lambda \Xi_s ds $ with $\Xi_s=
g(s,X_s) R^{-1} g(s,X_s)'$ and
\beaa
p_u(\tau)&=&p_0(\tau) \exp\left(-\int_{t}^T ds \frac{1}{2\lambda}u(s,X_s)'g(s,X_s)'\Xi^{-1}_s g(s,X_s)u(s,X_s)\right.\\
&+&\left. \int_{t}^T  \frac{1}{\lambda}u(s,X_s)' g(s,X_s)'\Xi_s^{-1} (dX_s-f(s,X_s)ds)\right)\\
&=&p_0(\tau)\exp\left(\frac{1}{\lambda}\left(\int_t^T ds \frac{1}{2}u(s,X(s))'Ru(s,X_s)+\int_t^T u(s,X(s))' R dW(s)\right)\right)
\eeaa
}
\bea
p_u(\tau)&=&\lim_{ds\to 0}\prod_{s=t}^{T-ds} \cN(X_{s+ds}|\mu_s,\Xi_s)\label{ce4a}\\
&=&p_0(\tau) \exp\left(-\int_{t}^T ds \frac{1}{2}u^2(s,X_s)+\int_{t}^T  u(s,X_s) g(s,X_s)^{-1} (dX_s-f(s,X_s)ds)\right)\nonumber
\eea
$p_0(\tau)$ is the distribution over trajectories in the absence of control, which we call the uncontrolled dynamics.
From this we obtain the Radon-Nikodym derivative
\bea
\frac{dp_0(\tau)}{dp_u(\tau)}&=&\exp\left(\sum_{s=t}^T ds \frac{1}{2}u^2(s,X_s)-\sum_{s=t}^T  u(s,X_s) g(s,X_s)^{-1} (dX_s-f(s,X_s)ds)\right)\nonumber\\
&=&\exp\left(-\int_t^T ds \frac{1}{2}u^2(s,X(s))-\int_t^T u(s,X(s)) dW(s)\right)\label{ce4}
\eea
where in the last step we used  dynamics Eq.~\ref{importance}. 
Using Eq.~\ref{ce4a} one immediately sees that
\beaa
\int d\tau p_u(\tau) \log \frac{p_u(\tau)}{p_0(\tau)}=\E_u \int_t^T  ds \frac{1}{2} u(s,X(s))^2
\eeaa
In other words, the quadratic control cost in the path integral control problem Eq.~\ref{controlproblem1} can be expressed as a KL divergence between the distribution over trajectories under control $u$ and the distribution over trajectories under the uncontrolled dynamics. 
 Eq.~\ref{controlproblem1} can thus be written as
\bea
J(t,x)&=& \min_u \int d\tau p_u(\tau) \left( \log \frac{p_u(\tau)}{p_0(\tau)}+V(\tau)\right)\label{ce8}
\eea
with $V(\tau)=\Phi(X_T)+\int_t^T ds V(s,X(s))$. Since there is a one-to-one correspondence between $u$ and $p_u$, one can replace the minimization with respect to the functions $u$ in Eq.~\ref{ce8} by a minimisation with respect to the distribution $p$ subject to a normalization constraint $\int d\tau p(\tau)=1$. 
The optimal solution is given by 
\bea
p^*(\tau)=\frac{1}{\psi(t,x)} p_0(\tau)\exp(-V(\tau)) \label{ce7}
\eea
where $\psi(t,x)=\E_{p_0} e^{-V(\tau)}$ is the normalization, which is identical to Eq.~\ref{eq1}. Substituting $p^*$ in Eq.~\ref{ce8} yields the familiar result $J(t,x)=-\log \psi(t,x)$.

Eq.~\ref{ce7} expresses $p^*$ in terms of the uncontrolled dynamics $p_0$ and the control cost. It suggests a Monte Carlo sampling scheme that samples from $p_0$ and weights with $e^{-V}$. From Eq.~\ref{ce4}, we can equivalently express Eq.~\ref{ce7} using importance sampling with importance sampling control $u$ as 
\bea
p^*(\tau)=\frac{1}{\psi(t,x)} p_u(\tau)\frac{dp_0(\tau)}{dp_u(\tau)}\exp(-V(\tau))=\frac{1}{\psi(t,x)} p_u(\tau) \exp(-S(t,x,u)) \label{ce9}
\eea
\subsection{The cross entropy method for path integral control}
\label{section:ce+pi}

We are now in a similar situation as the cross entropy method. We cannot compute the optimal control $u^*$ that parametrizes the optimal distribution $p^*=p_{u^*}$ and instead wish to compute a near optimal control $\hat{u}$ such that $p_{\hat{u}}$ is close to $p^*$. Following the CE argument, we minimise
\bea
KL(p^*|p_{\hat{u}})&\propto &-\E_{p^*}\log p_{\hat{u}}\label{ce10} \\
&\propto &\lim_{ds\to 0} \E_{p^*} \left(\sum_{s=t}^T \frac{1}{2} \hat{u}^2(s,X_s) ds-\hat{u}(s,X_s)g(s,X_s)^{-1}(X_{s+ds}-X_s-f(s,X_s)ds)\right)\nonumber \\
&=&\frac{1}{\psi(t,x)}\E_{p}e^{-S(t,x,u)} \int_t^T ds \left(\frac{1}{2}\hat{u}(s,X(s))^2-\hat{u}(s,X(s))\left(u(s,X(s))+\frac{dW_s}{ds}\right)\right)\nonumber
\eea
where in the second line we used Eq.~\ref{ce4} and discard the constant term $\E_{p^*} \log p_0$ and in the third line we used Eq.~\ref{ce9} to express the expectation with respect to the optimal distribution $p^*$ controlled by $u^*$ in terms of a weighted expectation with respect to an arbitrary distribution $p$ controlled by $u$. We further used that $X_{s+ds}= X_s+f(s,X_s)ds+g(s,X_s)(u(s,X_s)+dW(s))$. 
The expectation of $dW_s$ in Eq.~\ref{ce10} is non-zero due to the weighting by $e^{-S(t,x,u)}$. 
\footnote{For the special case of $p=p^*$ we have $e^{-S(t,x,u^*)}=\psi(t,x)$ and the $dW_s$ term vanishes. }

The $KL$ divergence Eq.~\ref{ce10} must be optimized with respect to
the functions $\hat{u}_{t:T}=\{\hat{u}(s,X_s), t\le s\le T\}$. In
addition, the $KL$ divergence involves an expectation value that uses
a sampling control $u_{t:T}=\{u(s,X_s), t\le s\le T\}$. We are free to
choose any sampling control as they all are unbiased estimators, but the more the sampling control resembles the optimal control, the more efficient can these expecations values be estimated. 

We now assume that $\hat{u}$ is a parametrized function with parameters $\theta$. In the time-dependent case, we consider different $\theta_s$ for each of the functions $\hat{u}(s,x|\theta_s)$ separately. 
In this case the gradient of the $KL$ divergence Eq.~\ref{ce4} is given by:
\bea
\frac{\partial KL(p^*|\hat{p})}{\partial \theta_s}&=&
\frac{1}{\psi(t,x)}\E_p e^{-S(t,x,u)}  \left(\hat{u}(s,X(s))-u(s,X(s))-\frac{dW_s}{ds}\right)\frac{\partial \hat{u}(s,X(s))}{\partial \theta_s}
\label{ce13}
\eea

In the case that $\hat{u}(s,x)$ and $u(s,x)$ are linear combinations of a set of $K$ basis functions $h_{sk}(x)$ with parameters $\theta_{sk}$ and $\theta^0_{sk}$, respectively,  ie. $\hat{u}(s,x)=\sum_{k=1}^K \theta_{sk} h_{sk}(x)$ and similar for $u(t,x)$, we can set the gradient equal to zero and obtain the set of equations:
\bea
\sum_{l=1}^K  \left(\theta_{sl}-\theta^0_{sl}\right) \av{h_{sl}h_{sk}}=\av{\frac{dW_s}{ds}h_{sk}}\qquad t\le s\le T,\quad k=1,\ldots, K\label{ce11}
\eea
where we defined $\av{F}=\frac{1}{\psi(t,x)}\E_p e^{-S(t,x,u)}  F$ with $p$ a distribution over trajectories under control $u$ that is linearly parametrized by $\theta^0$. Eq.~\ref{ce11} is for each $s$ a system of $K$  linear equations with $K$ unknowns $\theta_{sk}, k=1,\ldots,K$. The statistics $\av{h_{sl}h_{sk}}$ and $\av{\frac{dW_s}{ds}h_{sk}}$ can be estimated for all times $t\le s\le T$ simultaneously from a single Monte Carlo sampling run using the control $u$ parametrized by $\theta^0$. The fixed point equations Eq.~\ref{ce11} were derived in \cite{thijssen2014a} using a different reasoning.

Although in principle the optimal control explicitly depends on time, there may be reasons to compute a control function $\hat{u}(x)$ that does not explicitly depend on time. For instance, consider a stabilizing task such as an inverted pendulum. The optimal control solution $u^*(t,x)$ assumes an optimal timing of the execution of the swing-up. If for some reason this is not the case and the timing is off, an inappropriate control $\hat{u}(t,x)$ is used at time $t$. Another situation where a  time-independent solution is preferred is when the horizon time is very large, and the dynamics and the cost are also not explicit functions of time. The advantage of a time-independent control solution is clearly that it requires less storage. 

We thus consider $\hat{u}(X_s)$ and $u(X_s)$ independent of time parametrised by $\theta$ and $\theta^0$, respectively. 
In this case the gradient of the $KL$ divergence Eq.~\ref{ce10} is given by:
\bea
\frac{\partial KL(p^*|\hat{p})}{\partial \theta}&=&\frac{1}{\psi(t,x)}\E_{p}e^{-S(t,x,u)} \left(\int_t^T ds\left(\hat{u}(X(s))-u(X(s))\right)\frac{\partial \hat{u}(X(s))}{\partial \theta}\right.\nonumber\\
&-&\left.\int_t^T dW(s) \frac{\partial \hat{u}(X(s))}{\partial \theta}\right)
\label{ce14}
\eea
Note the extra integral over $s$, due to the fact that a single control function is active at all times. In the last term, the integration over $s$ has resulted in a It\^{o} stochastic integral. This has removed the awkward numerical estimation of $\av{\E dW(s)/ds}$. 

In the case that $\hat{u}(x)$ and $u(x)$ are linear combinations of a set of $K$ basis functions $h_{k}(x)$ with parameters $\theta_{k}$ and $\theta^0_{k}$, respectively, we can again set the gradient equal to zero and obtain the set of equations:
\bea
\sum_{l=1}^K  \left(\theta_{l}-\theta_{l}^0\right)\av{\int_t^T ds h_l(X(s)) h_k(X(s))}=\av{\int_t^T dW_s h_{k}(X(s))}\qquad k=1,\ldots, K\label{ce12}
\eea
Eq.~\ref{ce12} is a system of $K$  linear equations with $K$ unknowns $\theta_{k}, k=1,\ldots,K$. 

If required, the estimations of $\theta$ in Eqs.~\ref{ce11} and~\ref{ce12} can be repeated several times, each time with an improved $\theta,u$, implementing an adaptive importance sampling algorithm. In iteration $n$, $\theta=\theta_{n+1}$ is computed using a sampling control parametrized by $\theta^0=\theta_n$. 

In the case that $\hat{u}$ does not depend linearly on $\theta$ one
cannot directly solve $\frac{\partial KL(p^*|\hat{p})}{\partial
\theta}=0$. In this case one must resort to a gradient descent
procedure. In this case, one can also include the idea of adaptive
importance sampling. Remember that the $KL$ divergence Eq.~\ref{ce10}
must be minimized with respect to $\theta$ but also involves a
sampling control, parametrized by $\theta^0$. Since the gradient
descent procedure presumably monotonically improves the control, it is
best to use the most recent control estimate as sampling control.
Setting $u=\hat{u}$ in the gradients for the time-dependent and
time-independent cases Eqs.~\ref{ce13} and ~\ref{ce14} significantly
simplifies them and the gradient descent updates become
\bea
\theta_{s,n+1}&=&\theta_{s,n} - \eta \frac{\partial KL(p^*|\hat{p})}{\partial \theta_{s,n}}\big|_{u=\hat{u}_n}=\theta_{s,n} 
+\eta  \av{\frac{dW_s}{ds}\frac{\partial \hat{u}(s,X(s))}{\partial \theta_{s,n}}}\label{ce15}\\
\theta_{n+1}&=&\theta_{n} - \eta \frac{\partial KL(p^*|\hat{p})}{\partial \theta_{n}}\big|_{u=\hat{u}_n}=\theta_{n} 
+\eta \av{\int_t^T dW_s \frac{\partial \hat{u}(X(s))}{\partial \theta_{n}}}\label{ce16}
\eea
respectively,
and $\eta>0$ a small parameter. Since, Eqs.~\ref{ce15} and~\ref{ce16} are the gradients of the KL divergence, their convergence is guaranteed using standard arguments. We refer to this gradient method as the Path Integral Cross Entropy method or PICE.

\section{Numerical illustration}\label{section:numerical}

In this section, we illustrate path integral learning for two simple problems. For a linear quadratic control problem, where we compare the result with the optimal solution, and for an inverted pendulum control task where we compute the non-linear state feedback controller. 

\begin{figure}
\bc
\includegraphics[width=0.8\textwidth]{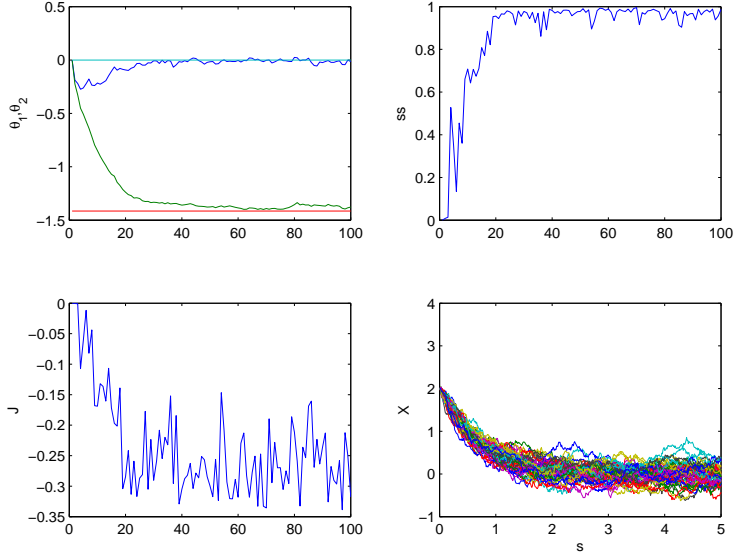}
\ec
\caption{Illustration of PICE Eq.~\ref{ce16} for a $1$-dimensional linear quadratic control problem with $Q=2, R=1, \nu=0.1, T=5$. We used time discretization $ds=0.01$ and generated $50$ sample trajectories for each gradient computation all starting from $x=2$ and $\eta=0.1$. 
The  top left plot shows $\theta_{1,2}$ as a function of gradient desent step. Top right shows effective sample size as a function of gradient descent step. Bottom left shows optimal cost to go $J$ as a function of gradient descent step. Bottom right shows 50 sample trajectories in the last gradeint descent iteration.}
\label{main_singlewell1}
\end{figure}

Consider the finite horizon $1$-dimensional linear quadratic control problem with dynamics  and cost 
\beaa
dX(s)&=& u(s,X(s))ds+dW(s)\qquad 0\le s\le T\\
C&=&\E \int_0^T ds \frac{R}{2}u^2(s,X(s)) +\frac{Q}{2} X(s)^2
\eeaa
with $\E dW(s)^2 =\nu ds$. The optimal control solution can be shown to be a linear feed-back controller
\beaa
u^*(s,x)&=&-R^{-1}P(s)x\qquad P(s)=
\sqrt{QR} \tanh
\left(\sqrt{\frac{Q}{R}}(T-s)\right)
\eeaa
For finite horizon, the optimal control explicitly depends on time, but  for large $T$ the optimal control becomes independent of $t$: $u^*(x)=-\sqrt{\frac{Q}{R}} x$. We estimate a time-independent feed-back controller of the form 
$\hat{u}(x)=\theta_1 + \theta_2 x$ using path integral learning rule Eq.~\ref{ce16}. The result is shown in fig.~\ref{main_singlewell1}.

Note, that $\theta_1,\theta_2$ rapidly approach their optimal values $0,-1.41$ (red and blue line). Under- estimation of $|\theta_1|$ is due to the finite horizon and the transient behavior induced by the initial value of $X_0$, as can be checked by initializing $X_0$ from the stationary optimally controlled distribution around zero (results not shown). 
The top right plot shows the entropic sample size defined as  the scaled entropy of the distribution: $ss=-\frac{1}{\log N}\sum_{i=1}^N \hat{w}_i \log \hat{w}_i$ and $\hat{w}_i=w_i/\hat{\psi}$ from Eq.~\ref{mc_psi1}, as a function of gradient desent step, which increases due to the improved sampling control.

As a second illustration we consider a simple inverted pendulum, that satisfies the dynamics
\[
\ddot{\alpha}= -\cos \alpha + u
\]
where $\alpha$ is the angle that the pendulum makes with the horizontal, $\alpha=3\pi/2$ is the initial 'down' position and $\alpha=\pi/2$ is the target 'up' position, 
$-\cos \alpha$ is the force acting on the pendulum due to gravity. 
Introducing $x_1=\alpha,x_2=\dot{\alpha}$ and adding noise, we write this system as
\beaa
dX_i(s)&=&f_i(X(s)) ds + g_i(u(s,X(s)+dW(s))\qquad 0\le s\le T, \quad i=1,2\\
f_1(x)&=&x_2\\
f_2(x)&=& -\cos x_1\\
g&=&(0,1)\\
C&=&\E \int_0^T ds \frac{R}{2}u(s,X(s))^2+\frac{Q_1}{2}(\sin X_1(s) -1)^2 + \frac{Q_2}{2}X_2(s)^2 
\eeaa
with $\E dW_s^2=\nu ds$ and $\nu$ the noise variance.

\begin{figure}
\bc
\includegraphics[width=0.3\textwidth]{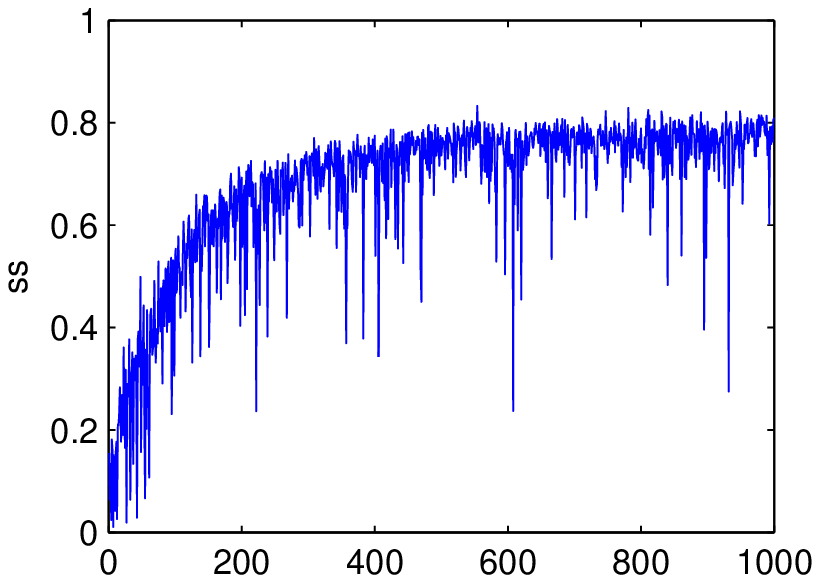}
\includegraphics[width=0.3\textwidth]{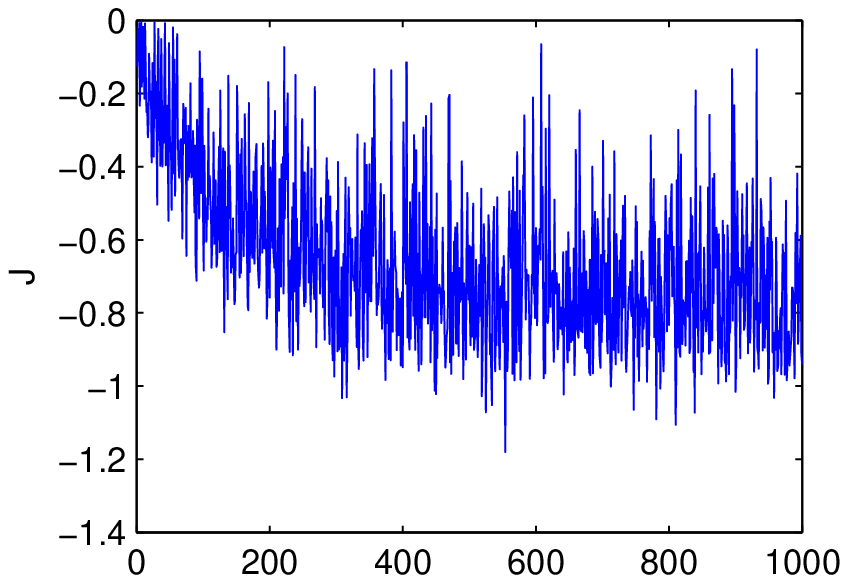}
\includegraphics[width=0.3\textwidth]{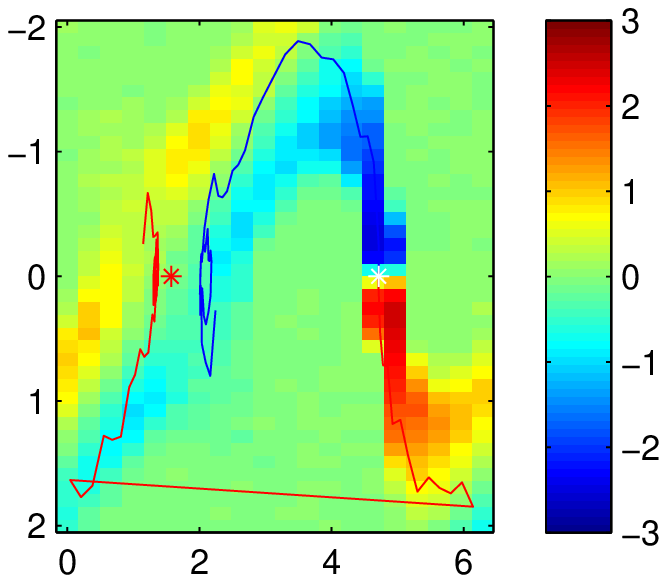}
\ec
\caption{Illustration of gradient descent learning Eq.~\ref{ce16} for a second order inverted pendulum problem with $Q_1=2, Q_2=0.02, R=1, \nu=0.3, T=5$. We used time discretization $ds=0.1$ and generated $500$ sample trajectories for each gradient computation all starting from $(x_1,x_2)=(-\pi/2,0)\pm(0,0.02)$ and $\eta=0.4$, $K_1=20,K_2=40$.
Left: Entropic sample size versus importance sampling iteration.
Middle: Optimal cost to go versus importance sampling iteration.
Right: Optimal control solution $\hat{u}(x_1,x_2)$ versus $x_1,x_2$ with $0\le x_1\le 2\pi$ and $-2\le x_2\le 2$. }
\label{main_pendulum2}
\end{figure}

We estimate a time-independent feed-back controller on a grid $k_1=1:K_1, k_2=1:K_2$, 
\beaa
\hat{u}(x_1,x_2)=\theta_{k_1,k_2} \qquad 
x^-_i+(k_i-1)dx_i \le x_i \le x^-_i+k_i dx_i,\quad i=1,2
\eeaa
with $x_i^\pm$ the maximum and minimum value of $x_i$ and $dx_i=(x_i^+-x_i^-)/K_i$. 
The results of the path integral learning rule 
Eq.~\ref{ce16}
are shown in fig.~\ref{main_pendulum2}.
Fig.~\ref{main_pendulum2}Left shows that the effective sample size increases with importance sampling iteration and stabalizes 
to approximately 80 \%.
Fig.~\ref{main_pendulum2}Middle shows the optimal cost-to-go decreases with importance sampling iteration. The fluctuation are due to
the finite constant learning rate $\eta$ and the sampling approximation of the expectation value in the gradient computation.
Fig.~\ref{main_pendulum2}Right shows the solution after 1000 importance sampling iterations in the $(x_1,x_2)$ plane. White star is initial location $(3\pi/2,0)$ (pendulum pointing down, zero velocity) and red star is the target state $x=(\pi/2,0)$ (pendulum point up, zero velocity). There are two example trajectories shown. The red trajectory forces the particle with positive velocity towards the top, and the blue solution forces the particle with negative velocity towards the top. Note the green NE-SW ridge in the control solution around the top. These are states where the position deviates from the top position, but with a velocity directed towards the top. So in these states no control is required. In the orthogonal NW-SE direction, control is needed to balance the particle.
This example shows that the learned state feedback controller is able to swing-up and 
stabilize the inverted pendulum.

It should be noted that the use of the path integral method for stabilizing stochastic control task is challenging, as is evident from the large fluctuations despite the large number of samples for these relatively small problems. 
The reasons are the following.
\bi
\item
The weights of the trajectories are proportional to $e^{-S}$ with $S\propto 1/\lambda$ from Eq.~\ref{ndim} and $\lambda = R \nu$ playing the role of temperature. Small
$\lambda$ has the effect that the effective sample size is small (close to one sample), because the weight of one trajectory dominates all other trajectories. Thus, in order to have a large effective number of samples one cannot choose $\nu$ too small, meaning that the stochastic disturbances will be relatively large which make the problem harder to control. In order to control these, the control should be sufficiently large, meaning that $R$ should be small. But $R$ cannot be chosen too small either since it affects the effective sample size in the same way as $\nu$. This problem is due to the log transform that is used to linearize the Bellman equation.
\item
No matter how complex or unstable the problem, if the control solution approaches the optimal control sufficiently close, the effective sample size should reach 100 \%. 
Representing the optimal control solution exactly requires in general an infinitely large model, except in special cases where a finite dimensional representation of the optimal control is known.
An infinite model requires infinitely many samples to avoid overfitting. Less than maximal entropic sample size is thus also due to 
the finite dimensionality of the model.
\ei
This suggests that the key issue for the succesful application of the path integral method is the parametrization that is used to represent $\hat{u}$. This representation should balance the two conflicting requirements of any learning problem: 1) the parametrization should be sufficiently flexible to represent an arbitrary function and 2) the number of parameters should be not too large so that the function can be learned with not too many samples.

The inverted pendulum can of course also be controlled using other methods, for instance using the iterative LQG. One first solves  the deterministic control problem in the absence of noise and then computes a linear feedback controller around this solution. In that case the solution is 'unimodal', representing one of the two possible swing-up solutions, and time-dependent. The point of the simulation is to illustrate that it is in principle possible to learn any state feedback controller, such as the 'multi-modal' control solution that represents both solutions simultaneously.

\section{Bayesian system identification: potential for neuroscience data analysis }
\label{neural}
We have shown that the path integral control problem is equivalent
to a statistical estimation problem. We can use this identity to solve
large stochastic optimal control problems by Monte Carlo sampling.
We can accelerate this computation by importance sampling and have
shown that the optimal control coincides with the optimal importance
sampler. In this section, we consider the reverse connection between
control and sampling: We consider the problem to compute the posterior
distribution of a latent state model that we wish to approximate using
Monte Carlo sampling, and to use optimal controls to accelerate this
sampling problem. 

In neuroscience, there is great interest for scalable inference methods,
e.g. to estimate network connectivity from data or decoding of
neural recordings. It is common to assume that there is an underlying
physical process of hidden states that evolves over time, which is
observed through noisy measurements. In order to extract information
about the processes giving rise to these observation, or to estimate
model parameters, one needs knowledge of the posterior distributions
over these processes given the observations. For instance, in the
case of calcium imaging, one can indirectly observe the network activity
of a large population of neurons. Here, the hidden states represent
the activity of individual neurons, and the observations are the calcium
measurements, {[}Mishchenko et al., 2011{]}. 

The state-of-the-art is to use one of many variations of particle
filtering-smoothing methods to estimate the state distributions conditioned
on the observations, see {[}Briers et al., 2010, Doucet and Johansen,
2011, Lindsten and Schoen, 2013{]}. A fundamental shortcoming of these
methods is that the estimated smoothing distribution relies heavily
on the filtering distribution which is computed using particle filtering.
For high dimensional problems these distributions may differ significantly
which yields poor estimation accuracy, as seen in the following
example.

One can easily see that the path integral control computation is mathematically
equivalent to a Bayesian inference problem in a time series model
with $p_{0}(\tau)$ the distribution over trajectories under the forward
model Eq. 1 with $u=0$, and where one interprets $e^{-V(\tau)}=\prod_{s=t}^{T}p(y_{s}|x_{s})$
as the likelihood of the trajectory $\tau=x_{t:T}$ under some fictitious
observation model $p(y_{s}|x_{s})=e^{-V(x_{s})}$ with given observations
$y_{t:T}$. The posterior is then given by $p^{*}(\tau)$ in Eq. 21.
One can generalize this by replacing the fixed initial state $x$
by a prior distribution over the initial state. Therefore, the optimal
control and importance sampling results of section 3.2 can be directly
applied. The advantage of the PI method is that the computation scales
linear in the number of particles%
\footnote{The details of this approach are left for a following paper.%
}, compared to the state-of-the-art particle smoother that scales quadratic
in the number of particles, although in practice significant accelerations
can be made, e.g. \cite{fearnhead2010sequential,lindsten2013backward}. 

\begin{figure}
\bc
\includegraphics[width=0.9\textwidth]{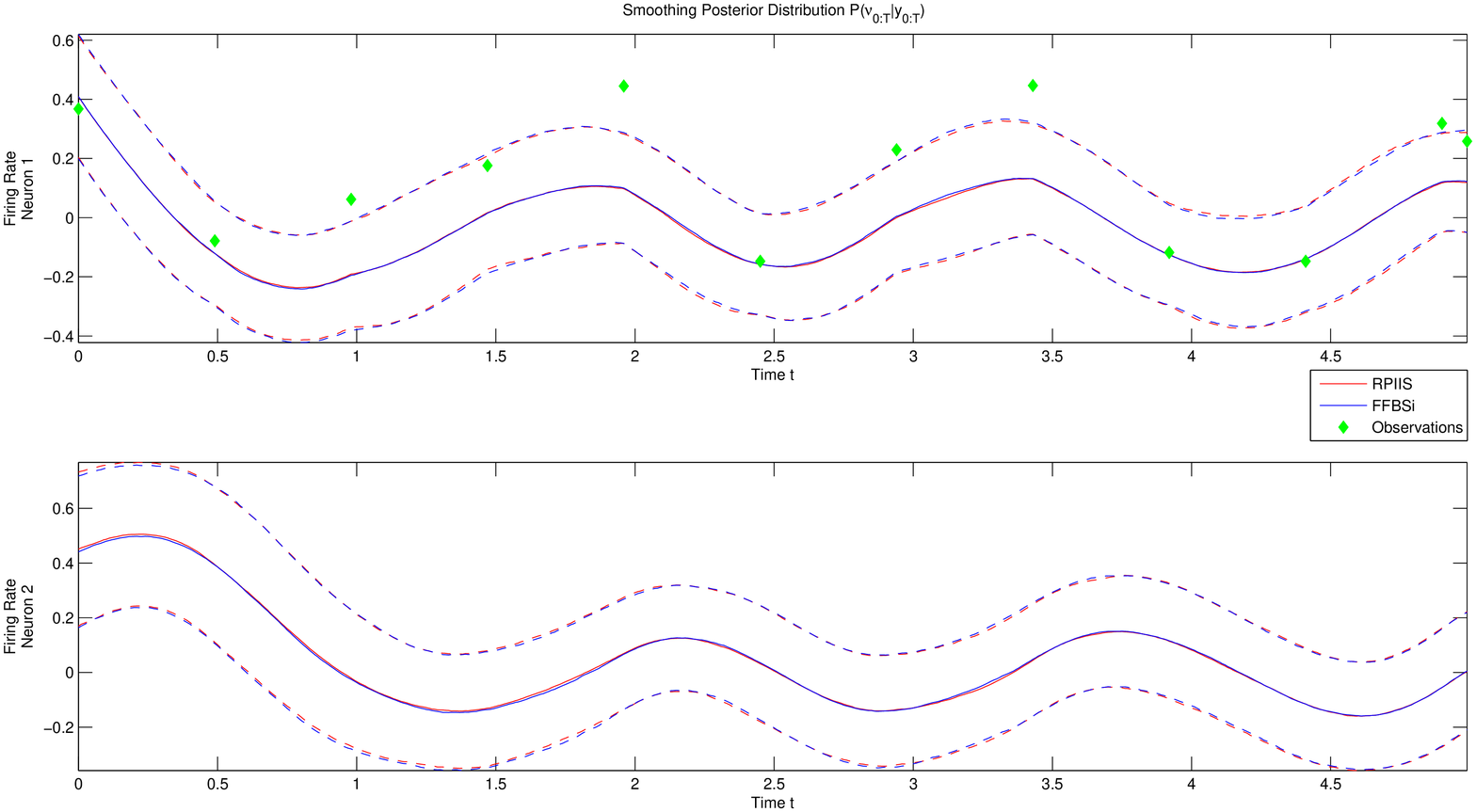}
\ec
\caption{Comparison of path integral control (RPIIS) and the forward filter backward smoother (FFBSi cf.  {[}Lindsten and Schoen, 2013{]}) for a 2-dimensional neural network, showing mean and one standard deviation of the marginal posterior solution for both methods.} 
\label{fig:SDestim}
\end{figure}

To illustrate this we estimate the posterior distribution of a noisy
2-dimensional firing rate model given 12 noisy observations of a single
neuron, say $\nu_{1}$ (green diamonds
in fig.~\ref{fig:SDestim}). The model is given by 
\[
\frac{d\nu_t}{dt}=-\nu_{t}+\tanh(J*\nu_{t}+\theta)+\sigma_{dyn}dW_{t}
\]
$J$ is a $2$-dimensional antisymmetric matrix and $\theta$ is a $2$-dimensional vector, both 
with random entries from a Gaussian distribution with
mean zero and standard deviation 25 and standard deviation 0.75, respectively, and  $\sigma_{dyn}^{2}=0.2$. 
We assume a Gaussian observation model $\cN(y_i|\nu_{1t_i},\sigma^2_\mathrm{obs})$ with $\sigma_\mathrm{obs}=0.2$.
We generate the 12 $1$-dimensional observations $y_i,i=1,\ldots, 12$ 
with $\nu_{1t_i}$ the firing rate of neuron 1 at time $t_i$ during one particular run of the model.

We parametrized the control as $u(x,t)=A(t)x+b(t)$
and estimated the 2x2 matrix $A(t)$ and the $2$-dimensional vector
$b(t)$ as described in {[}Thijssen and Kappen, 2015{]} and Eq.~\ref{ce11}. 

The path integral control solution (RPIIS) is shown in fig.~\ref{fig:SDestim}) and
was computed using 22 importance sampling iterations with 6000 particles
per iteration. 
As a comparison, the forward-backward particle filter
solution (FFBSi) was computed using N = 6000 forward and M = 3600
backward particles. In blue, we see the FFBSi estimates and in red
the RPIIS estimates of the posterior distribution $p(\nu_{0:T}|y_{0:T})$. The computation time was 35.1 s and 638 s respectively. 

\begin{figure}
\bc
\includegraphics[width=0.9\textwidth]{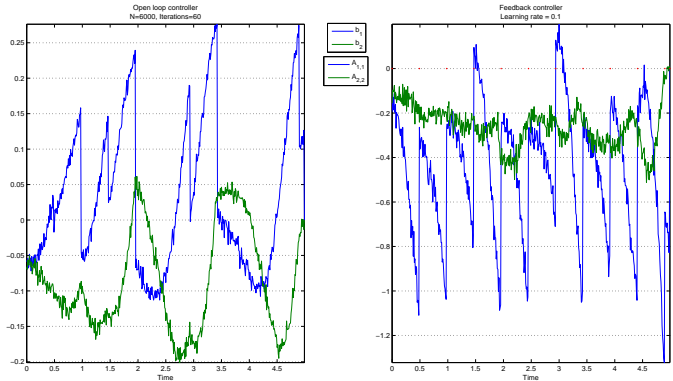}
\ec
\caption{Control parameters; Left: Open-loop controller $b_i(t),i=1,2$; Right: Diagonal entries of feedback linear controller $A_{ii}(t),t=1,2$}
\label{fig:Control}
\end{figure}

Figure ~\ref{fig:Control} shows the estimated control parameters used for
the RPIIS method. 
The open loop controller $b_1(t)$ steers the particles to the observations. The feedback controller $A_{11}(t)$ 'stabilizes' the particles around the observations (blue lines). Due to the coupling between the neurons, the non-observed neuron is also controlled in a non-trivial way. 
To appreciate the effect of using a feedback controller, we compared these results with an open-loop controller $u(x,t)=b(t)$. 
This reduces the ESS from 60 \% for the feedback controller to around 29 \% for the open loop controller. The lower
sampling efficiency increases the error of the estimations, especially the variance of the posterior marginal (not shown).
When choosing the importance sampling controller, there is in general a trade off between accuracy and the computational effort involved in the update rules in Eqs.~\ref{ce11} or~\ref{ce12}.


The example shows the potential of adaptive importance sampling for
posterior estimation in continuous state-space models. A publication with the analysis of this
approach for high dimensional problems is in preparation. This can be used to accelerate maximum
likelihood based methods to estimate, for instance connectivity, decoding of
neural populations, estimation of spike rate functions and, in general,
any inference problem in the context of state-space models; see \cite[ and references therein]{oweiss2010statistical} for a treatment of state-space models in the contex of neuroscience and neuro-engineering.

\section{Summary and discussion}
The original path integral control result of Theorem~\ref{thm1} expresses the optimal control $u^*(t,x)$ for a specific $t,x$ as a Feynman-Kac path integral. 
The important advantage of the path integral control setting is that, asymptotically, the result of the sampling procedure does not depend on the choice of sampling control. The reason is that the control used during exploration is an importance sampling in the sense of Monte Carlo sampling and any importance sampling strategy gives the same result asymptotically. Clearly, the efficiency of the sampling depends critically on the sampling control. 
Theorem~\ref{thm1} can be used very effectively for high dimensional stochastic control problems using
the Model Predictive Control setting \cite{gomez2015real}. 

However, Theorem~\ref{thm1} is of limited use when we wish to compute a parametrized control function for all $t,x$. We have therefore here proposed the cross entropy argument, originally formulated to optimize importance sampling distributions, to find a control function whose distribution over trajectories is closest to the optimally controlled distribution. In essence, this optimization replaces the original KL divergence $KL(p|p^*)$ Eq.~\ref{ce8} by the reverse KL divergence $KL(p^*|p)$ and optimizes for $p$. The resulting path integral learning method provides a flexible framework for learning a large class of non-linear stochastic optimal control problems with a control that is an arbitrary function of state and parameters.
The idea to optimize this reverse KL divergence was earlier explored for the time-dependent case and linear feedback control in \cite{gomez2014}. 
%

We have restricted our numerical examples to parametrizations that are linear in the parameters. Generalization to non-linear parametrizations, such as for instance (deep) neural networks, Gaussian processes or other machine learning methods can be readily considered, at no significant extra computational cost. 

\cite{de2005tutorial} also discuss the application of the CE method to a Markov decision problem (MDP), which is a discrete state-action control problem. 
The main differences with the current paper are that we discuss the continuous state-action case. Secondly, the MDP problem is formulated as an optimization problem to find $x^*=\mathrm{argmax}_x  f(x)$. \cite{de2005tutorial} provide a generic approach to apply the CE method to optimization, by defining a distribution $p(x)$ and optimise the expected cost  $C=\sum_x p(x) f(x)$ with respect to $p$. By construction, the optimal $p$ is of the form $p(x)=\delta_{x,x^*}$, ie. a distribution that has all its probability mass on the optimal state
\footnote{Generalizations restrict $p$ to a parametrized family $p(x|\theta)$ and optimize with respect to $\theta$ instead of $p$ direction \cite{mannor2003cross}.}. The CE optimization computes this optimal zero entropy/zero temperature solution starting from an initial random (high entropy/high temperatue) solution. As a result of this implicit annealing, it has been reported that the CE method applied to optimization suffers from severe local minima problems \cite{szita2006learning}. An important difference for the path integral control problems that we discussed in the present paper is the presence of the entropy term $p(x) \log p(x)$ in the cost objective. As a result, the optimal $p$ is a finite temperature solution that is not peaked at a single state but has finite entropy. Therefore, problems with local minima are expected to be less severe.

The path integral learning rule Eq.~\ref{ce16} has some similarity with the so-called policy gradient method for average reward reinforcement learning \cite{sutton1999policy}
\beaa
\Delta \theta=\eta \E_\pi  \sum_a \frac{\partial \pi(a|s)}{\partial \theta}Q^\pi(s,a)
\eeaa
where $s,a$ are discrete states and actions, $\pi(a|s,\theta)$ is the policy which is the probability to choose action $a$ in state $s$, and $\theta$ parametrizes the policy. $\E_\pi$ denotes expectation with respect to the invariant distribution over states when using policy $\pi$ and $Q^{\pi}$ is the state-action value function (cost-to-go) using policy $\pi$. 
The convergence of the policy gradient rule is proven when the policy is an arbitrary function of the parameters. 

The similarities between policy gradient and path integral learning are that the policy takes the role of the sampling control and the policy gradient involves an expectation with respect to the invariant distribution under the current policy, similar to the time integral in Eq.~\ref{ce16} for large $T$ when the system is ergodic. The differences are 1) that the expectation value in the policy gradient is weighted by $Q^\pi$, which must be estimated independently, whereas the brackets in Eq.~\ref{ce16} involve a weighting with $e^{-S}$ which is readily available; 2) Eq.~\ref{ce16} involves an It\^{o} stochastic integral whereas the policy gradient does not; 3) the policy gradient method is for discrete state and actions and the path integral learning is for controlled non-linear diffusion processes; 4) the policy gradient expectation value is not independent of $\pi$ as is the case for the path integral gradients Eqs.~\ref{ce13} and~\ref{ce14}.

\section{Acknowledgement}
I would ike to thank Vicen{\c{c}} G{\'o}mez for helpful comments and careful reading of the manuscript. 

\appendix
\section{It\^o calculus}
Given two diffusion processes, 
\begin{align}
dY &= A(Y)ds + B(Y)dW\label{y} \\
dZ &= C(Z)ds + D(Z)dW\nonumber
\end{align}
the It\^o's product rule gives the evolution of the product process
\begin{align}
d(YZ) 
	&= YdZ + ZdY + d[Y,Z] \nonumber\\
d[Y,Z] &= B(Y)D(Z) ds \label{ito_product}
\end{align}
The term in the last line is known as the quadratic covariance.

Let $F(Y)$ as a function of the stochastic process $Y$.
It\^o's Lemma is a type of chain rule that gives the evolution of $F$;
\bea
dF=dY \partial_y F + \frac{1}{2} d[Y,Y] \partial_y^2 F=\left(A\partial_y F +\frac{1}{2}B^2\partial_y^2 F\right)ds +B\partial_y F dW \label{ito_lemma}
\eea

Putting a process Eq.~\ref{y} in integral notation and taking the expected value yields the following
\begin{align}
Y
	&= \int Ads+ \int B dW \label{ito_integral}\\
\E[ Y ]
	&= \int \E[A] ds 
\end{align}
\\ \\
The It\^o Isometry states that
\begin{align}
\E\left[ \int A(Y)dW \int B(Y)dW  \right]
	&= \int \E[A(Y)B(Y)] ds \label{ito_iso}
\end{align}

\small{ 
\bibliography{/Users/bertkappen/doc/authors}
\bibliographystyle{apalike} }

\end{document}